# Topological charge localization in bilayer graphene induced by an antisymmetric electric potential step

J. C. Martinez[1], M. B. A. Jalil

*Information Storage Materials Laboratory, Electrical and Computer Engineering Department, National University of Singapore, 4 Engineering Drive 3, Singapore 117576*

S. G. Tan

*Data Storage Institute, DSI Building, 5 Engineering Drive 1, National University of Singapore, Singapore 117608*

A charged particle whose energy is less than the electric potential step it is incident upon, is expected to undergo partial reflection and transmission. In bilayer graphene, however, a potential step in the form of an antisymmetric kink results in particle localization due to the interaction between the particle and its chiral partner. It is found that when the potential step exceeds a threshold, zero-energy modes of the system emerge, and causes the kink to acquire a charge. The Hall-effect plateaus in the vicinity of the zero modes correspond, unexpectedly, to those of the monolayer. The topological nature of these kink-induced effects and the ease with which a kink can be generated in practice, suggest possible applications in e.g. storage of information or switching devices.

---

[1] elemjc@nus.edu.sg

One can enjoy bowling in two ways: a normal person can hurl the ball and be thrilled by strikes and spares, while a child can gently slide the ball on the sidetrack and watch it roll back from the return rack. Another possibility occurs if the ball neither goes for the pins nor comes back but stays put somewhere. Paradoxically this scenario can be more intriguing as one tries to analyze the resting position of the ball. An analogous situation occurs in quantum mechanical tunneling, where electrons are either transmitted or reflected if their kinetic energies are greater or less than the strength of the step-potential barrier they are incident upon [1]. But in relativistic quantum mechanics, a third scenario can theoretically arise, namely, spatially-bound particles due to the interaction between particle and antiparticle. Such interplay is unknown in semi-classical or non-relativistic quantum mechanics, and is rarely observed even in relativistic quantum mechanics since the requisite electric fields far exceed available technologies. In this paper, we will show that this effect can be reproduced in bilayer graphene in the presence of an antisymmetric potential kink.

The phenomenon we describe requires a particle-antiparticle pair to be held apart by an external electrostatic field and yet be strongly correlated. For Dirac electrons the typical energy is $O(mc^2)$ and the corresponding correlation length is the Compton wavelength $\hbar/mc$ [2], implying a restraining electric field of the order of $10^{17}\,\mathrm{V\,m^{-1}}$, a field far beyond present-day capabilities. But we may use bilayer graphene whose top speed is 300 times smaller than $c$ and the excitation mass one-twentieth of the electron's [3]. (Hence chiral pairs replace particle-antiparticle pairs.) Then the electric field needed is 10 orders of magnitude smaller, or $10^7\,\mathrm{V\,m^{-1}}$, which is accessible with present technology. Monolayer graphene under these circumstances would generate a different mechanism, more akin to the Zener breakdown, and quite unrelated to our purpose [4, 5].

We therefore study a gated bilayer graphene configuration with an impressed voltage kink $V$ to provide a restraining potential for a particle and its chiral partner. Such kinks can be produced for instance in a graphene p-n junction [6]. Recently it has been shown that such configurations can support zero modes and chiral states in the vicinity of the domain wall separating the insulating regions [7]. If the bias $V(x)$ in the form of a kink is applied between the layers then the motion in the $y$-direction is that of a free particle and the dynamics in the $x$-direction will be the one of interest. We will show that a charged particle with energy less than the bias will not undergo total reflection as expected quantum mechanically but will remain in the vicinity of the kink and can manifest itself as charge bound to the kink. Once the kink has surpassed a threshold strength, the system is able to support zero modes and the Hall conductivity plateaus correspond to those of the graphene monolayer [3]. All these imply that the kink introduces new and unexpected features into the bilayer dynamics which are externally adjustable. We expect our general results to be of interest in other non-graphene areas of investigation, e.g., particle physics [8] and superconductivity [9].

The low-energy Hamiltonian $\tilde{H}$ for the graphene bilayer is a $4 \times 4$ matrix in the space spanned by the four-component wave function ($\psi_{A1}$, $\psi_{B1}$, $\psi_{A2}$, $\psi_{B2}$), where the subscript letter and number label the inequivalent atoms (A or B) in the layer (1 or 2). It can be reduced further into a reasonably accurate effective $2 \times 2$ matrix if we are interested only in the lowest-energy bands, i.e., when the interlayer hopping between nearest neighbors is much larger than the electron energy (measured from zero momentum). If we model the kink potential as an anti-symmetric tanh profile imposed by the electrostatic bias, the effective $2 \times 2$ wave equation is [7, 10]

$$\begin{pmatrix} -(e + r \tanh x) & \left(\frac{d}{dx} + p_y\right)^2 \\ \left(\frac{d}{dx} - p_y\right)^2 & -(e - r \tanh x) \end{pmatrix} \begin{pmatrix} u(x) \\ v(x) \end{pmatrix} = 0, \qquad (1)$$

where $r$ ($0 \leq r \leq 1$) denotes the bias strength and $p_y$ the particle momentum in the $y$-direction. This effective Hamiltonian involves the atoms $(A_2, B_1)$, which are not linked by the interlayer hoppings. We will refer to the equations obtained from Eq. (1) as the first and second component equations, respectively. Formally Eq. (1) can be reduced into a single equation because it is clear that one component converts to the other through the replacement $x \to -x$. Thus, a possible set of solutions to Eq. (1) can be obtained, in which their two components are related by $u(x) = \pm v(-x)$. Eq. (1) has been treated as two separate problems, one for $\Psi = (u(x), u(-x))^T$ and another for $\Phi = (u(x), -u(-x))^T$ [7]. However, we will not adopt the above approach, as the resulting solutions would appear to imply a *nonlocal* relation between $u(x)$ and $u(-x)$ for the *entire* range of $x$. But in this system there is no underlying symmetry to support such a relation. Instead, we follow an alternative method below, in which a *local* relation will suffice to obtain an exact solution.

To solve Eq. (1) exactly we consider the ansatz:

$$\begin{pmatrix} u \\ v \end{pmatrix} = e^{(\alpha+i\beta)x} \begin{pmatrix} U(x) \\ V(x) \end{pmatrix}, \qquad (2)$$

where $\alpha$ and $\beta$ are real constants and replace $x$ with the auxiliary variable $z = -e^{-2x}$. We are principally interested in the intragap solutions, $|e| \leq r$, and we give explicit expressions for $r = 1$. We have for the first component of Eq. (1):

$$(1-z)zV'' + (1-z)(1-p_y+\alpha+i\beta)V' - \tfrac{1}{2}U = 0, \qquad (3)$$

where ' denotes $d/dx$. In arriving at (3) we had imposed the eigenvalue relation $(p-\alpha-i\beta)^2 V - (e+1)U = 0$, whose consistency will be verified later. Then we are led to consider two possibilities: $U(x) = V(x)$ and $U(x) = -V(x)$. With either choice, Eq. (3) is a hypergeometric equation with the solutions [11]

$$V(x) = \pm U(x) = \pm {}_2F_1(a,\ b,\ c,\ z), \qquad (4)$$

$$a = \tfrac{1}{2}\left(\sqrt{\pm(e-1)} + \sqrt{\pm(e+1)}\right),\ b = -\tfrac{1}{2}\left(\sqrt{\pm(e-1)} - \sqrt{\pm(e+1)}\right),\ c = 1+a+b,$$

in which ${}_2F_1(a,\ b,\ c,\ z)$ is the Gauss hypergeometric series and the parameters $a$, $b$, $c$ depend only on the energy. The eigenvalue relation can now be recast without reference to $U$ and $V$ as: $(p_y - \alpha - i\beta)^2 \mp (e+1) = 0$. This last result shows that our putative eigenvalue condition in Eq. (2) is consistent: it has no dependence on $x$, $U$, or $V$. We can express the solutions as: $e^{(\alpha+i\beta)x}\begin{pmatrix}V(x)\\V(x)\end{pmatrix}\equiv g_+(x)\begin{pmatrix}1\\1\end{pmatrix}$ and $e^{(\alpha+i\beta)x}\begin{pmatrix}-V(x)\\V(x)\end{pmatrix}\equiv h_-(x)\begin{pmatrix}-1\\1\end{pmatrix}$.

The second component can be solved similarly and we find $V(x)=\pm U(x)=\pm {}_2F_1(a,-b,1+a-b,z)$ with the corresponding eigenvalue relation $(p_y+\alpha+i\beta)^2 \mp (e-1)=0$. Similarly, this second set can be cast as $h_+(x)\begin{pmatrix}1\\1\end{pmatrix}$ and $g_-(x)\begin{pmatrix}-1\\1\end{pmatrix}$. Since both component equations describe the same system we combine the eigenvalue relations into a single equation:

$$((\alpha+i\beta)^2 - p_y^2)^2 = e^2 - 1, \qquad (5)$$

with the solutions $\alpha(\beta)=\frac{1}{\sqrt{2}}[(p_y^4+1-e^2)^{1/2}+(-)p_y^2]^{1/2}$.

We found two solutions of Eq. (1) depending on the sign of the relation $V(x)=\pm U(x)$, with the requisite global eigenvalue conditions which are independent of $x$ [see e.g. Eq. (5)]. We now describe how these solutions can be written explicitly. For the first component of Eq. (1) we have: for $x > 0$, $g_{+>}(x)=$
$=e^{-\alpha x}{}_2F_1(a,b,c,z)(c_1 e^{i\beta x}+c_2 e^{-i\beta x})$, and $c_1$ and $c_2$ are complex constants; a similar expression, $g_{+<}(x)$, can be given for $x < 0$ with the sign of $x$ reversed and with the coefficients $c_3$ and $c_4$ instead. These wave functions are normalizable, and the four coefficients $c_i$ are determined by demanding that $g_{+>}(x), g_{+<}(x)$ and their first three derivatives be continuous at $x = 0$. These yield four homogeneous simultaneous equations and we use the requirement that the determinant of the system must vanish to derive the dispersion relation, $e = e(p_y)$ [7]. Clearly, the only independent dynamical parameter is the energy, $e$. The same procedure is repeated for the wave function $h_-(x)$ (and similarly $h_+(x)$ and $g_-$), thus yielding a pair of complete solutions of the first (second) component of Eq. (1).

We now address how the solutions of the first and second components of Eq. (1) are related, which is pertinent to resolving the question of nonlocality raised earlier. The procedure above produced two pairs but there is really only one pair of solutions. Examining the solutions, $g_+(x)\begin{pmatrix}1\\1\end{pmatrix}$, etc and, in particular, the expressions for the parameters *a, b, c*, etc, we observe that $g_+(x)$ can be obtained from $g_-(x)$ by reversing the sign of the energy (and vice versa). The wave functions $h_+(x)$ and $h_-(x)$ are similarly connected. Hence the solutions of the two components are really chiral conjugates of the each other (i.e., $g \leftrightarrow h$) with $\begin{pmatrix}1\\1\end{pmatrix}$ transforming into $\begin{pmatrix}-1\\1\end{pmatrix}$, and vice versa. This can be shown formally by transforming Eq. (1) into its chiral conjugate form, wherein the sign of the energy *and* bias are reversed. It follows that $\begin{pmatrix}u\\v\end{pmatrix}$ is conjugate to $-\sigma_z\begin{pmatrix}u\\v\end{pmatrix}$, consistent with the results just obtained. If we recall that the original graphene Hamiltonian was derived within the tight-binding model [12, 13] and view the system in terms of a set of coupled oscillators, then the states $\begin{pmatrix}1\\1\end{pmatrix}$ and $\begin{pmatrix}-1\\1\end{pmatrix}$ just represent the two distinct and independent normal (symmetric and antisymmetric) modes of the system.

There is no need to invoke a nonlocal relation to solve Eq. (1). That the two sets of solutions pertain to only one (chiral) pair of solutions is implicit in Eq. (5) where we had combined the eigenvalue conditions for the two component equations (1): indeed $\alpha$ and $\beta$ must be obtained from Eq. (5) and not from their original defining equations.

In Figs. 1 and 2 we have plotted the dispersion relations. Starting from a bias with strength $r = 0$, and increasing it to $r = 1$, the two initial ($r = 0$) parabolic bands of the bilayer are gradually separated by the biasing potential (Fig. 1 a, b) creating a gap between the bands. Each band is further split into two sub-branches, this separation becoming obvious when the potential is stronger (Fig. 1 c). Eventually for a sufficiently large $r$ (Fig. 2), i.e. when the kink voltage exceeds a certain bias threshold, the lower (upper) sub-branch of the positive (negative) band is pulled far enough down (up) and crosses the $e$-axis, and creates zero-energy modes. It is interesting to note that these branches cross each other at zero energy and fixed $y$-momentum $\pm p_{0y}$, so that we have a gapless half-conical structure resembling the band-structure for the graphene monolayer. Overall, the dispersion graph corresponding to $u(x) = -v(-x)$ (Fig. 2) delineates the outline of a cat's head, while that corresponding to $u = v(-x)$ forms the reflected image of the former.

We can show that the Hamiltonian (1) reduces to the monolayer case in the vicinity of the zero mode, i.e., $|e| \ll 1$. Take for definiteness $U = V$ and assume $U = f\,g$, $f$ being the zero-mode solution. $g$ is a slowly varying function of $e$ which has a value of 1 at $e = 0$, and varies over a length scale which is much larger than the kink width.

Substituting $U$ into the Eq. (1) and making use of the eigenvalue relations we find that the Hamiltonian for $g$ is $\begin{pmatrix} 0 & ip_x + p_y \\ -ip_x + p_y & 0 \end{pmatrix}$ correct to $O(e^2)$, which is of the form $\vec{\sigma} \cdot \vec{p}_\perp$, where $\vec{p}_\perp$ is perpendicular to the zero-mode momentum. Thus, at the vicinity of the zero mode (around $e = 0$), the Hamiltonian mimics that of monolayer graphene, with its energy spectrum having the characteristic $\sqrt{N}$ signature [3, 14 - 16]. At the zero mode the dispersion curve is almost vertical, so we find reason to assert that $\vec{p}_\perp$ lies in the positive $p_x \geq 0$ half-plane of momentum space (see Fig. 2). For monolayer graphene, its momentum domain is the entire momentum space. In the present case, however, particles are constrained from taking negative values so that their paths do not enclose the origin.

As emphasized by Martin *et al.*, the zero modes of our system are not Dirac fermions but chiral modes specific to the bilayer. The fact that these modes occur only when the bias is sufficiently strong indicates that a topological kink is *insufficient* of itself to generate such modes. That a kink is necessary is clear because the decay of wave functions as $x \to \pm\infty$ requires the chiral pair to be close to each other near the origin. Each particle of the pair in turn is held in place by this electrostatic bias (with opposite signs on both sides of $x = 0$) along with the interaction with its conjugate. We can check the consistency of the above from a computation of the topological charge of the Fermi point. Writing the Hamiltonian (1) as $\vec{\phi}(\vec{p}, x) \cdot \vec{\sigma}$, this topological charge is given by [17]

$$N_3 = \frac{1}{8\pi} \varepsilon^{abc} \varepsilon_{ijk} \int_\Sigma dS^k \hat{\phi}^a \partial_{p_i} \hat{\phi}^b \partial_{p_j} \hat{\phi}^c \tag{6}$$

where $\Sigma$ is a surface enclosing the origin of the $p$-plane and may be taken to be two infinite planes parallel to the $p_x - p_y$ plane, one to the right and the other to the left of the origin (the separation between the planes being infinitesimal). This charge gives the difference between the number of right-moving and left-moving zero modes. Then, $N_3 = 0$, since we must sum (not subtract) the contributions from the right side and left sides of $\Sigma$. This is clearly consistent with our results.

Graphs of the real part of the wave functions are shown in Fig. 3. All the intragap states are bound and symmetric about the origin, a consequence of the local relation between $u$ and $v$, unlike Martin, *et al.* [7, 18]. We can also understand qualitatively why a minimum bias is required before zero modes appear. In the introduction, we noted that our system requires a strong correlation between the chiral pairs, which implies that they were within, say, a lattice spacing apart. The decay length of the wave function is $\alpha^{-1}$ (cf. Eq. (5) and replace 1 by $r^2$). For fixed $p_y$ and potential strength, $r$, this length is smallest when $e = 0$; on the other hand for fixed $p_y$ and $e$, this decay length is smaller as $r$ is increased. Thus $r$ has to be large enough for zero modes to appear. This trend is seen in Fig. 3 where we observe that the zero mode width is about half that of the dashed curve.

Because any negative energy eigenstate is related to a positive energy eigenstate by a unitary transformation, the local density of states $\rho(r) = \Sigma_e \psi_e^+(r) \psi_{e'}(r) \delta(e - e')$ is symmetric about $e = 0$ and the negative and positive energy eigenstates contribute equally. Including the zero modes, the conservation of the total number of states implies that the difference in densities with and without the kink $\delta\rho$ is

$$0 = \int d^2 r \left( 2\int_{-\infty}^{0^-} \delta\rho(r,e)de + |\psi_0(r)|^2 \right) \tag{7}$$

The two zero modes are normalized, so the integral of $\delta\rho$ over energy and space is – 1. Taking electron spin into account this means a charge of – $2e$ and total spin zero for the valence band (and conduction band). If the zero modes are unoccupied, the charge and spin for the kink are, respectively, $Q = -2e$ and $S = 0$; if the zero modes are singly (doubly) occupied, then $Q = -e$ (0), $S = ½$ (0). These serve as the signature of the presence of charge and the confinement of the chiral pair in the vicinity of the kink. Since the zero modes occur in pairs we do not see charge fractionalization here [19]. Moreover, we need not be concerned here with a violation of Kramer's theorem [20].

The zero modes may find possible application in two ways: (a) since the zero modes appear only beyond a threshold potential, their emergence can be used as a switching indicator and (b) in so far as zero modes are of two types associated with the chiral functions $\begin{pmatrix} 1 \\ 1 \end{pmatrix}$ and $\begin{pmatrix} -1 \\ 1 \end{pmatrix}$, they might store information much as binary bits do. The topological properties of the charge and spin of these zero modes confer a certain degree of robustness to these binary states. Further application of the zero modes can be derived from utilizing the valley degree of freedom (or "valleytronics" [21]), which can be modified along the kink direction. Our results would also be of interest in brane theory [8] and superconductivity [9].

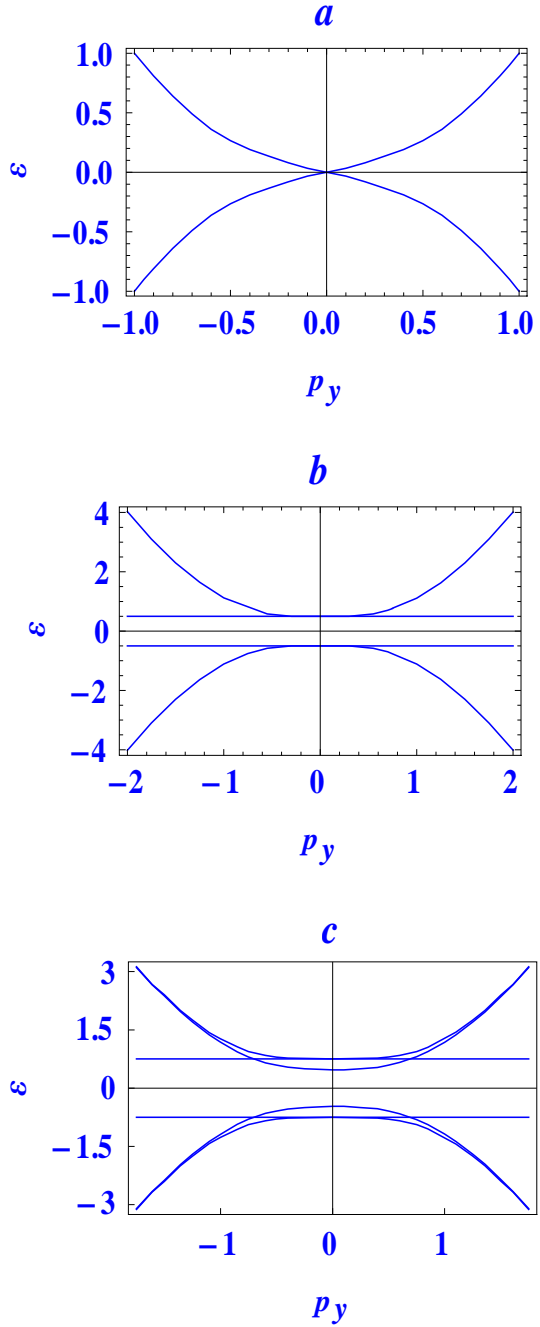

FIG. 1 Dispersion curves for (a) $r = 0$, (b) $r = 0.5$ and (c) $r = 0.75$. Graph (a) shows the characteristic pair of parabolic bands in the absence of a potential. For (b) there is hardly any splitting of the bands; (c) shows splitting into four bands but no zero modes.

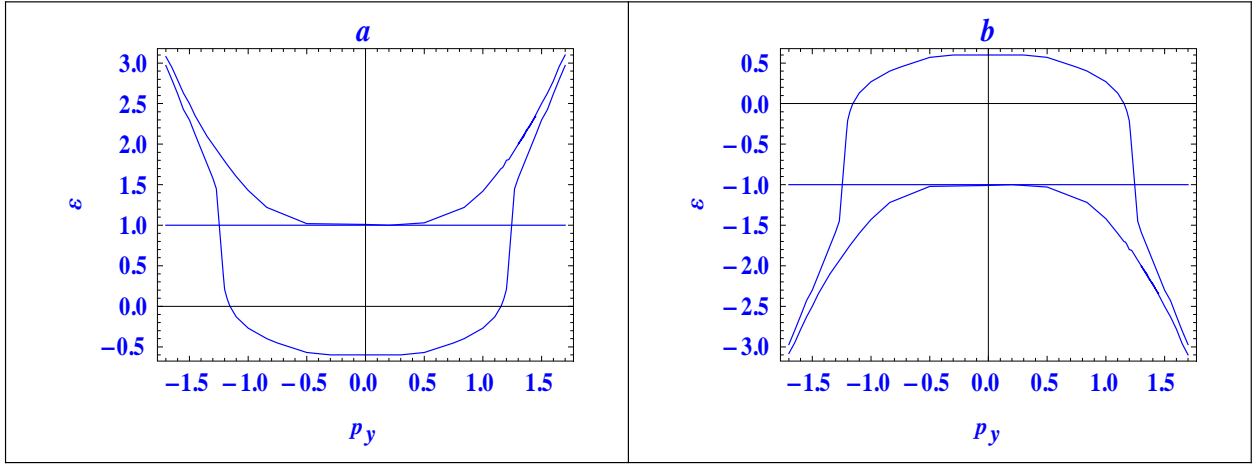

FIG. 2 Dispersion curves for *r* = 1. The graph (a) corresponds to *u* = -*v* whereas graph (b) to *u* = *v*. We observe zero modes at $p_{0y} = \pm 1.154$. When the two curves are superimposed we obtain a *half* conical structure around the zero modes because of the practically vertical line at $p_{0y}$.

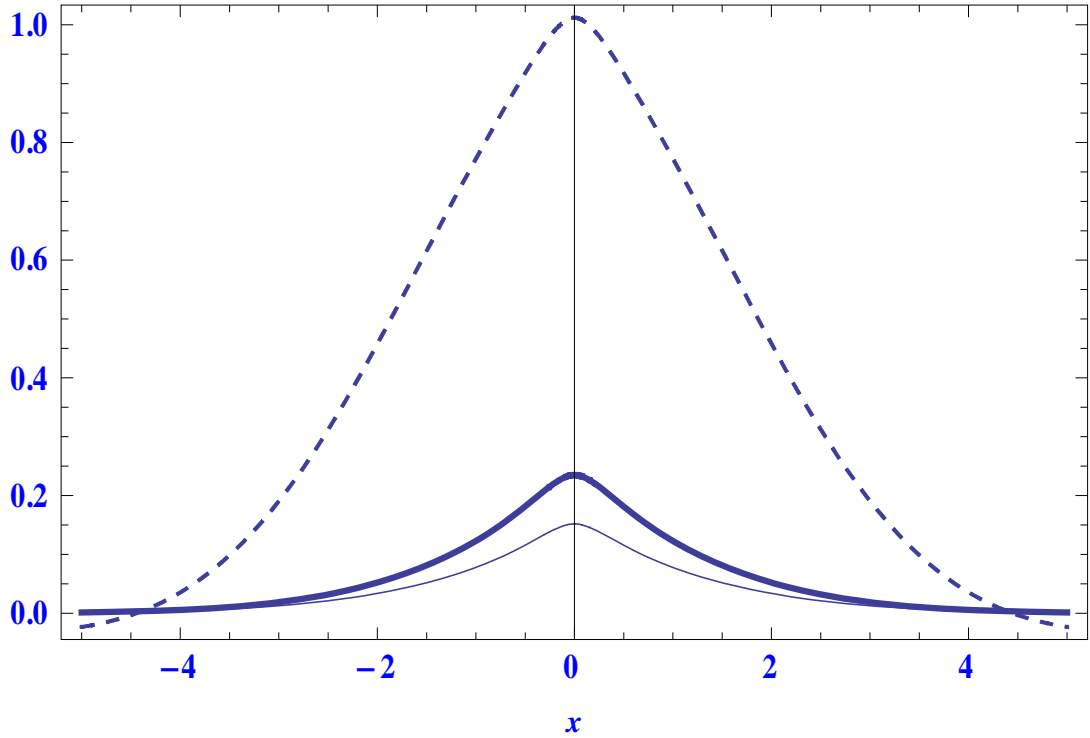

FIG. 3 Graphs of the real part of $u$ for three cases: $p_y = 0$, $e = -0.6$ (dashed), $p_y = 1.154$, $e = 0$ (thick), $p_y = 1.21$, $e = 0.55$ (thin). The thick graph corresponds to the zero mode. The length scale is in inverse lattice spacing; the vertical scale is the maximum value of the three.